\documentclass[aps,pre, twocolumn, floatfix, superscriptaddress, showpacs, showkeys]{revtex4-1}
\usepackage{epsf,amsmath,amssymb,verbatim,color,multirow,pifont,graphicx,cleveref,tabularx,cancel,soul}
\usepackage[dvipsnames]{xcolor}
\usepackage{rotating}

\setstcolor{red}

\crefname{equation}{Eq.}{Eqs.}
\crefname{section}{Sec.}{Secs.}
\crefname{figure}{Fig.}{Figs.}
\crefname{table}{Table}{Tables.}

\newcommand{\DSLrev}[1]{#1}
\newcommand{\DSLfinal}[1]{#1}

\begin{document}

\title{Recent advances of percolation theory in complex networks}
\author{Deokjae Lee}
\affiliation{CCSS, CTP and Department of Physics and Astronomy, Seoul National University, Seoul 08826, Korea}
\author{Y. S. Cho}
\affiliation{Department of Physics, Chonbuk National University, Jeonju 54896, Korea}
\author{K.-I. Goh}
\affiliation{Department of Physics, Korea University, Seoul 02841, Korea}
\author{D.-S. Lee}
\affiliation{Department of Physics, Inha University, Incheon 22212, Korea}
\author{B. Kahng}
\email{bkahng@snu.ac.kr}
\affiliation{CCSS, CTP and Department of Physics and Astronomy, Seoul National University, Seoul 08826, Korea}
\date{\today}

\begin{abstract}
During the past two decades, percolation has long served as a basic paradigm for network resilience, community formation and so on in complex systems. While the percolation transition is known as one of the most robust continuous transitions, the percolation transitions occurring in complex systems are often of different types such as discontinuous, hybrid, and infinite-order phase transitions. Thus, percolation has received considerable attention in network science community. Here we present a very brief review of percolation theory recently developed, which includes those types of phase transitions,  critical phenomena, and finite-size scaling theory. Moreover, we discuss potential applications of theoretical results and several open questions including universal behaviors.      
\end{abstract}

\maketitle

\section{Introduction}
What are complex systems? A complex system consists of many constituents which generate emerging behaviors through diverse interactions~\cite{complex1,complex2}. One of the powerful ways of examining the intrinsic nature of a complex system is to observe how such emerging patterns change by the small perturbation applied to the system. In complex systems, such a change or response is so sensitive to the details of the perturbation that it is extremely diverse. In such a case, it is not adequate to predict definitely how much the change would be. Recently, Parisi argued~\cite{parisi} that the prediction for the responses to small perturbations in complex systems cannot be made definitely but can be in a probabilistic way. He showed examples of protein structures in biological systems and spin glasses in physical systems. In the case of proteins, subject to small external perturbations such as the change in pH or the substitution of a single amino acid, they would fold to a completely different 3D structure but with practically the same free energy. In the case of the disordered magnetic systems, each spin responds to a slowly varying external field by changing its orientation, forming a series of bursts, known as Barkhausen noise~\cite{barkhausen}. The number of spin bursts depends on the disorder strength of the system, following a power-law distribution at a critical strength of disorder. The prediction of the number of spin bursts in this case can only be probabilistic. The stock market is another example of complex systems. Stock prices are determined as a result of the complicated interplay between numerous investors, and the price changes were also found to exhibit a power-law distribution~\cite{finance}. All these examples aptly illustrate how the concept of probabilistic prediction may apply as a new paradigm in modern science. Other examples can also be found in fields as diverse as meteorology and geology~\cite{buchanan}.

At the brink of the 21st century, two papers heralding the beginning of a new science of network, which were for small-world networks by Watts and Strogatz (WS)~\cite{ws} and for scale-free (SF) networks by Barab\'asi and Albert (BA)~\cite{ba99}, were published. Since then, complex network research has flourished, not only as an active research field but also as a common platform on  which the systems approach to various complex systems in a wide variety of multidisciplinary fields such as the natural, social, and information sciences can be potentially unified~\cite{review1,review2,review3}.

A complex system is represented by a graph or network whose nodes and links stand for its constituents and interactions, respectively. In such complex networks, there is no regular structure and position, but nodes are connected randomly to other nodes. The origin of such a random graph was proposed by Erd\H{o}s and R\'enyi (ER) long ago~\cite{er}. Separation between nodes in a giant cluster on a macroscopic scale is counted by the number of hops across nodes, so-called Hamming distance. The average distance between two nodes is short, being logarithmically or double logarithmically proportional to the system size, i.e., the number of nodes in the system, denoted as $N$. Such complex networks are small-world in general term. 

Complex networks observed in the real world such as the World Wide Web~\cite{www} and Internet~\cite{Internet} are heterogeneous in their structures: A few nodes (called hubs) are connected with many other nodes and the rest many nodes are connected with a few other nodes. The number of connections, called degree, of each nodes form a power-law or a heavy-tailed distribution. The networks with a power-law degree distribution are called SF networks~\cite{ba99}. To check if the formation of SF-type networks is an  intrinsic characteristic in complex systems, we examined the diameter change by the deletion of a single node for various real-world SF networks. We found that the diameter changes are indeed diverse and their distribution exhibits a power-law decay with an exponent less than three. Thus, the variance of the distribution diverges as $N\to \infty$. Thus, we cannot predict definitely how much the diameter changes by the deletion of a single node, but do in a probabilistic way~\cite{diameter_change}. 

A percolation transition of complex networks is an appropriate instance of exhibiting structural complexity~\cite{resilience2,cohen00,redner,mendes_percolation,lee04}. At a percolation threshold, the cluster size distribution follows a power law with an exponent less than three. Thus, one cannot predict the mean cluster size definitely at the percolation threshold. Percolation transition is known as one of the most robust continuous transitions. However, diverse types of percolation transitions such as explosive, discontinuous, hybrid, and infinite-order phase transitions are observed in complex networks. Such diverse types of percolation transitions are shown schematically in Fig.~1. In this paper, we introduce properties of those percolation transitions briefly. 
       
This paper is organized as follows. We introduce in Sec.~\ref{sec:models} several models of complex networks and their main features. In Sec.~\ref{sec:percolation}, we briefly review the percolation theory on regular lattice and of complex networks. Next, we review several research works of the explosive percolation model in Sec.~\ref{sec:ep} and of the hybrid percolation transition in Secs.~\ref{sec:hybrid} and \ref{sec:mcc}.  Finally Sec.~\ref{sec:summary} is devoted to summary and discussion.

\begin{figure}
\centering
\includegraphics[width=.96\linewidth]{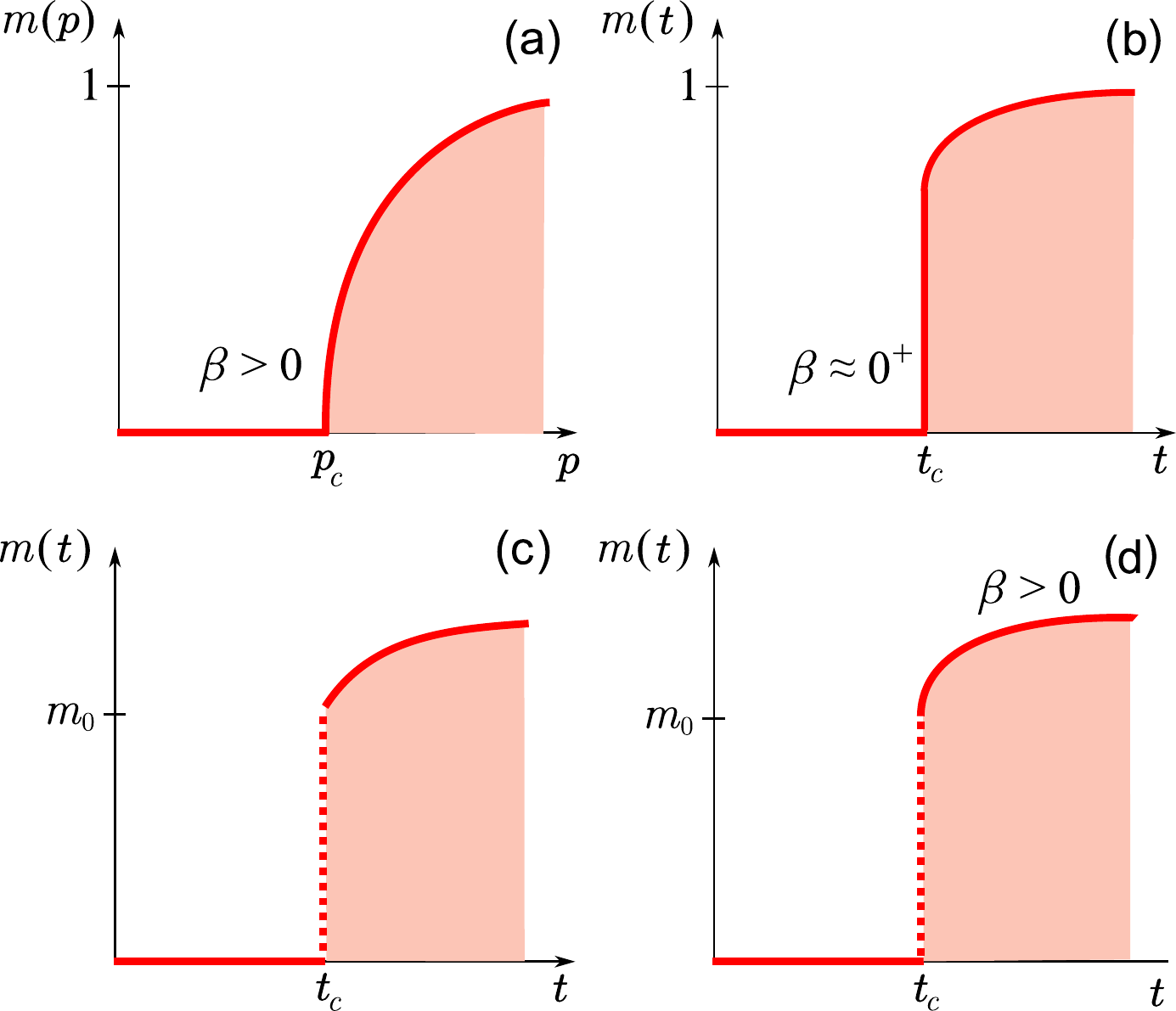}
\caption{
(Color online) Schematic figures of the order parameter $m$ for (a) continuous, (b) explosive, (c) discontinuous and (d) hybrid percolation transitions. For (b), the exponent $\beta$ is not zero but extremely small. For (c), $m(t)$ does not exhibit a critical behavior for $m > m_0$, where $m_0$ is the discontinuity of the order parameter. For (d), $m(t)$ exhibits a critical behavior for $m(t) > m_0$ with $\beta >0$. Dotted lines in (c) and (d) represent jumps, whereas solid vertical line in (b) does continuous increase.}
\label{fig1}
\end{figure}

\section{Network models}
\label{sec:models}

Complex networks may be classified into two types: random networks with Poissonian degree distribution and SF networks with power-law degree distribution. Here we introduce a few prototypical models for those types of networks. 

\subsection{The Erd\H{o}s and R\'{e}nyi model}
A simplest model for random networks was proposed by Erd\H{o}s and R\'{e}nyi (ER)~\cite{er}. In the ER model, $N$ nodes are present from the beginning and edges are added one by one between a pair of nodes selected randomly at each time step. Due to the random choices, the distribution of the number of edges incident on each node, called the degree distribution, follows the Poisson distribution,
\begin{equation}
P_d(k)=e^{-\langle k \rangle}\frac{\langle k \rangle^k}{k!},
\end{equation}
where $k$ denotes degree, the number of edges connected to a certain node and $\langle k \rangle$ is its average over all nodes. 

The mean distance between two nodes, denoted as $\langle \ell \rangle$, depends on the system size $N$ as  
\begin{equation}
\langle \ell \rangle \propto \frac{\ln N}{\ln \langle k \rangle}.
\end{equation}  
$\langle \ell \rangle$ of the ER network is shorter than that of hypercubic lattice in Euclidean space $\sim N^{1/d}$, where $d$ is spatial dimension.

\subsection{Scale-free network models}
It was revealed that many real-world networks such as the World-Wide Web, Internet, the coauthorship, the protein interaction network display power-law or heavy-tailed behaviors in the degree distribution. The power-law behavior is written as 
\begin{equation}
P_d(k)=(\lambda-1)k_{\rm min}^{\lambda-1} k^{-\lambda},
\label{eq:sf_degree_dist}
\end{equation}
where $k_{\rm min}$ is the smallest degree for which the power law holds. 
Such networks are called SF networks~\cite{ba99}. 
Thanks to recent extensive studies of SF networks, various properties of SF network structures have been uncovered~\cite{p_report}. 

For SF networks, the mean distance between two nodes is given as 
\begin{align}
\langle \ell \rangle=\left\{
\begin{array}{lll}
{\rm const.} \quad \quad & {\rm for}~~  &\lambda=2 \\[2mm] 
\ln\ln N \quad \quad &{\rm for}~~ &2 < \lambda <3 \\[2mm]
\frac{\ln N}{\ln \ln N} \quad \quad & {\rm for}~~ &\lambda=3 \\[2mm]
\ln N \quad \quad & {\rm for}~~ &\lambda > 3. 
\end{array}
\right.
\label{eq:diameter}
\end{align}
We remark that when $2 < \lambda < 3$, $\langle d \rangle$ increases with the system size $N$ as $\ln\ln N$, a significantly slower than the $\ln N$ derived for the ER random networks. Thus, SF networks with $2 < \lambda <3$ are called ultra-small networks~\cite{ultra_small}.  

How are SF network models constructed? Here we introduce several models. First, the configuration model was introduced by Molloy and Reed~\cite{molloy1,molloy2}. In this model, degree of each node is pre-assigned and represented by stubs or half-links, following a power law with the exponent $\lambda$, Eq.~\eqref{eq:sf_degree_dist}. Then two stubs are selected from different nodes and connected by an edge. This process is repeated until all stubs are connected. This model can produce a SF network without generating any degree-degree correlation. 

Secondly we introduce the static model~\cite{static_model}, which includes $N$ nodes from the beginning. Nodes are indexed by an integer $i$ ($i=1,\dots,N$) and assigned  the weight $w_i=i^{-\mu}$, where $\mu$ is a control parameter in [0,1). Next, we select a pair of nodes $(i,j)$ with probabilities $p_i\equiv w_i/\sum_k w_k$ and $p_j\equiv w_j/\sum_k w_k$, respectively, and add an edge between them unless they are already connected. This process is repeated until $L=NK$ edges are made in the system, where $K$ is a control parameter. Then the  mean degree becomes $2K$. However, because edges are connected to a vertex with frequency proportional to the weight of that vertex, the mean degree of vertex $i$ is given as 
\begin{equation}
\langle k_i \rangle=2L\frac{(1-\mu)}{N^{1-\mu}}i^{-\mu},
\end{equation}
Then it follows that the degree distribution follows the power law, $P_d(k)\sim k^{-\lambda}$, where $\lambda$ is given by $\lambda=1+1/\mu$. This model is useful for analytic calculations. However, this model generates degree-degree correlation when the degree exponent is in the range $2 < \lambda < 3$. 

To construct SF networks with no degree-degree correlation, Chung and Lu (CL) introduced a model~\cite{chung}, which is a simple modification of the static model. The CL model is similarly defined but the weight is given as $p_i \propto (i+i_0-1)^{-\mu}$, where $i_0\propto N^{1-1/2\mu}$.

\section{Percolation}\label{sec:percolation}

Percolation was first introduced in the 1950s to describe the flow of a fluid in a disordered medium~\cite{hammersley_1954}. However, the basic idea of percolation was effectively considered in the early 1940s in the study of gelation in polymers~\cite{flory1,flory2,flory3}. After those pioneering works, the concept of percolation was applied to a variety of natural and social phenomena and systems such as the spread of disease in a population~\cite{disease}, conductor--insulator composite materials~\cite{con_insul}, stochastic star formation in spiral galaxies~\cite{schulman_star_perc}, dilute magnets~\cite{dilute_magnet}, the resilience of systems~\cite{resilience1,resilience2,resilience3}, the formation of public opinion~\cite{opinion1,opinion2}, and nonvolatile memory chips~\cite{nonvolatile_memory_1,nonvolatile_memory_2}. Percolation has long served as a model for diverse phenomena and systems. The percolation transition, that is, the formation of a giant cluster on a macroscopic scale, is known as one of the most robust continuous transitions.\cite{stauffer,review1}. For instance, polymerization was modeled as percolation on the Bethe lattice~\cite{fisher1,fisher2}. 

Until recently, percolation has been studied mainly on regular lattices such as a square lattice in two dimensions. Each site (bond) on the square lattice is occupied by a conductor with probability $p$, which is a control parameter. Occupied sites at the nearest neighbors are regarded as connected, so current can flow between them if one site is charged. The system is located between two electrodes that are connected externally to a voltage source. As $p$ is increased, the connected conductors form a cluster. When $p$ is increased beyond a certain threshold $p_c$, the largest cluster can span the system, so pathways exist through which current can flow from the top to the bottom. Thus, $p_c$ is called a percolation threshold or a transition point. Unlike the case in spin models, the percolation transition is a geometric phase transition from an unconnected to a connected state. The fraction of occupied sites belonging to the spanning cluster becomes the order parameter of the percolation transition, which is denoted as $m(p)$. 

The order parameter $m(p)$, that is, the fraction of nodes belonging to the giant cluster, emerges at the percolation threshold $p_c$ and increases continuously from zero as the control parameter $p$ is increased beyond $p_c$. Near the percolation threshold, the order parameter exhibits critical behavior in the limit $N\to \infty$ as follows:
\begin{equation}
m(p)=\left\{
\begin{array}{lr}
0 & ~{\rm for}~~  p < p_c, \\
a(p-p_c)^{\beta} & ~{\rm for}~~ p \ge p_c, 
\end{array}
\right.
\label{eq:second_order}
\end{equation}
where $a$ is a constant, and $\beta$ is the critical exponent of the order parameter. 
The susceptibility is defined as $\chi_m \equiv \sum_s^{\prime} s^2 n_s(p)/\sum_s^{\prime} sn_s(p)$, where $n_s(p)$ is the number of clusters of size $s$ per $N$ at a certain point $p$ and the primed summations run over finite clusters. $\chi_m(p)$ diverges as $\chi_m \sim (p-p_c)^{-\gamma}$ in the thermodynamic limit $N\to \infty$ and behaves as $\chi_m \sim N^{\gamma/\bar{\nu}}$ at the transition point $p_c$ in finite systems, where $\bar{\nu}=d\nu$. 

The percolation transition can be represented in terms of a spin model using the formalism of the $q$-state Potts model of Kasteleyn and Fortuin~\cite{kasteleyn}. At $p_c$, the cluster sizes are very inhomogeneous. The size distribution of finite clusters behaves as $n_s(p)\sim s^{-\tau} e^{-s/s^*}$ for $p \ne p_c$, where $s^*$ is a characteristic cluster size and scales as 
\begin{align}
s^*\sim |p-p_c|^{-1/\sigma}.
\end{align} 
At $p=p_c$, $n_s(p)\sim s^{-\tau}$. Thus, the first and second moments of $n_s(p)$ become 
\begin{align}
&1-\sum_s^{\prime} s n_s \sim s^{*(2-\tau)}\sim |p-p_c|^{(\tau-2)/\sigma}, \\
&\sum_s^{\prime} s^2 n_s \sim s^{*(3-\tau)} \sim |p-p_c|^{-(3-\tau)/\sigma}, 
\label{sns}
\end{align}
respectively, where the primed summations go over finite clusters. 
Using the identity 
$$m(p)=1-\sum_s^{\prime} sn_s(p),$$ 
one can show that the singular behavior of the first and second moments of $n_s(p)$ becomes equivalent to $m(p)$ and $\chi(p)$, respectively. Thus, the critical exponents become 
\begin{align}
\beta=(\tau-2)/\sigma
\end{align} and 
\begin{align}
\gamma=(3-\tau)/\sigma,
\end{align} 
respectively.  

In percolation, the linear size of a typical cluster is the correlation length, denoted as $\xi$. For $p < p_c$, there are many finite clusters in the system. The total number of clusters per $N$ is given as $\sum_s n_s(p)$, which leads to $s^{*(1-\tau)}\sim (p-p_c)^{(\tau-1)/\sigma}$. On the other hand, there exist $N/\xi^d\sim N(p-p_c)^{\bar{\nu}}$ clusters in the system. Thus, one can obtain a hyperscaling relation $\bar{\nu} \sigma=(\tau-1)$. Similarly, one can obtain another hyperscaling relation, $\alpha+2\beta=\bar{\nu}$.

\subsection{Network formation} 

There have been a few attempts to describe scale-free networks in the framework of equilibrium statistical physics~\cite{berg02,manna03,burda,doro03,farkas}. In this approach, various mathematical tools developed in equilibrium statistical physics may be used to understand network structures. To proceed, one needs to define equilibrium network ensembles with appropriate weights, where one graph corresponds to one state of the ensemble. In a canonical ensemble, the number of links $L$ is fixed: Given a degree distribution, $P_d(k)$, the mean degree $\langle k\rangle \equiv \sum k P_d(k)$ is obtained. Then the number of links obtained through the relation, $L=\langle k\rangle N/2$, can be fixed. A degree sequence specifies the number of nodes with degree $k$ as $P_d(k)N$~\cite{molloy1,newman01}.

A grandcanonical ensemble can be also defined, where the number of links is a fluctuating variable while keeping the SF nature of the degree distributions. The grandcanonical ensemble for SF random graphs is realized in the static model introduced in Ref.~\cite{static_model} or in its generalized version investigated in Refs.~\cite{caldarelli02}. The name `static' originates from the fact that the number of nodes is fixed from the beginning. Here each node $i$ has a prescribed normalized weight $p_i$ and a link \DSLrev{is assigned to a pair of nodes $i$ and $j$ with probability} $p_ip_j$~\cite{soderberg02,chung,aiello02}. \DSLrev{After the edge attaching process is repeated $NK$ times with $K$ a parameter controlling the total number of links, distinct nodes $i$ and $j$ are connected by a link with probability $f_{ij}$  given as} 
\begin{equation}
f_{ij}=1-(1-2p_ip_j)^{NK}\approx
1-e^{-2NKp_ip_j}.
\label{fij}
\end{equation} 

\DSLrev{A graph can be defined by the adjacency matrix $\mathbf{A}$ the elements of which are $A_{ij}=1$ if node $i$ and $j$ are connected and $A_{ij}=0$ otherwise.}   Given the occupation probability $f_{ij}$, the probability that a graph \DSLrev{with an adjacency matrix $\mathbf{A}$} is formed is given as
\begin{eqnarray}
P_{\rm graph}(\mathbf{A})&=&\DSLrev{\prod_{i\ne j}  [ f_{ij} A_{ij} + (1-f_{ij})(1-A_{ij})]}
\nonumber \\ 
&=&e^{-NK(1-M_2)}\prod_{\DSLrev{(i,j), A_{ij}=1}} (e^{2NKp_ip_j}-1), 
\label{pg}
\end{eqnarray}
where $M_n\equiv \sum_{i=1}^N p_i^n$. 
The \DSLrev{expected value}  of a given quantity $O$  is obtained \DSLrev{by averaging over all possible adjacency matrices as}
\begin{equation}
\langle O \rangle =\sum_{\mathbf{A}} P_{\rm graph}(\mathbf{A}) \, O(\mathbf{A}).
\label{ensave}
\end{equation}

\DSLrev{If we use $p_i = i^{-\mu}/\sum_{j=1}^N j^{-\mu}$ with $0\leq \mu <1$ as the node selection probability in the static model, the ensemble of the obtained graphs finds the expected degree distribution to behave asymptotically as a power-law. To see this, we} define \DSLrev{the} generating function of degree $k_i$ of node $i$ as   $g_i(\omega)\equiv \langle \omega^{k_i}\rangle$. Then it becomes  
\begin{equation}
g_i(\omega)=\prod_{j(\ne i)} \left[
\DSLrev{1-f_{ij} + \omega f_{ij}}
\right],
\label{giz}
\end{equation}
where $f_{ij}$ is in Eq.~(\ref{fij}). Using this generating function, we can calculate the \DSLrev{expected} degree of node $i$ as 
\begin{equation}
\langle k_i \rangle = \left.\omega {d \over
	d\omega}g_i(\omega)\right|_{\omega=1}= \sum_{j(\ne i)} \DSLrev{f_{ij}
\simeq}  2KNp_i \propto i^{-\mu}, 
\label{kibn}
\end{equation}
and other quantities such as the mean degree as $\langle k \rangle=\sum_i \langle k_i \rangle /N \simeq 2K$. \DSLrev{The approximations are valid as long as $K\ll N^{1-\mu}$~\cite{lee04}.} Thus, the total number of links presenting in the system is determined stochastically with mean value $\langle L \rangle=\langle k \rangle N/2\simeq K$. 

\DSLrev{$g(\omega)=(1/N)\sum_i g_i(\omega)$ becomes the generating function of the degree distribution. The expansion of $g(\omega)$ around $\omega=1$  has a leading singular term $(1-\omega)^{1/\mu}$ in the region $k_{\rm max}^{-1} \ll1-\omega\ll 1$, which means that the  degree distribution $P_d(k)$ behaves asymptotically for $1\ll k\ll k_{\rm max}$ as}
\begin{equation}
P_d(k)
\simeq k^{-\lambda},
\label{degree}
\end{equation}
\DSLrev{where the degree exponent $\lambda$ is related to $\mu$ by $\lambda = 1+ {1\over \mu}$ and $k_{\rm max} =\langle k_1\rangle\sim N^{\mu}$~\cite{lee04}}. 

\subsection{Potts model formulation}
\label{sec:Potts}

\DSLrev{The $q$-state Potts model attracted much attention due to its rich and general critical behaviors, including the Ising model at $q=2$ and the Ashkin-Teller model at $q=4$~\cite{wu}. In particular,}
the $q\to 1$ limit of the Potts model corresponds to the bond percolation problem, \DSLrev{which can be applied to the study of percolation in random graphs~\cite{lee04}} \DSLfinal{as reviewed in this section.} 

Consider the $q$-state Potts Hamiltonian given by 
\begin{widetext}
\begin{eqnarray}
-\mathcal{H}=2NK \sum_{i>j} p_i p_j \delta_{\sigma_i,\sigma_j} + 
H \sum_{i=1}^N [q\delta_{\sigma_i,1}-1], 
\label{hm}
\end{eqnarray}
\end{widetext}
where $K$ is the interaction strength appearing in Eq.~(\ref{fij}), $H$ is an \DSLrev{external} field, \DSLrev{$\delta_{x,y}$ is the Kronecker delta function, and $\sigma_i=1,2,\ldots, q$ is the Potts spin. }
The partition function $Z_N(q,H)$ can be written as 
\begin{widetext}
\begin{eqnarray}
Z_N(q,H)={\rm Tr}e^{-\mathcal{H}} &=& {\rm Tr}\prod_{i>j} \left[
1+(e^{2NKP_iP_j}-1) \delta_{\sigma_i,\sigma_j}\right]\prod_i e^{H (q\delta_{\sigma_i,1}-1)} \nonumber \\
&=&e^{NK(1-M_2)} \sum_{\DSLrev{\mathbf{A}}} P_{\rm graph}(\DSLrev{\mathbf{A}}) 
\prod_{s\geq 1} (e^{srH}+re^{-sH})^{n_\mathbf{A}(s)},
\label{znqh}
\end{eqnarray}
\end{widetext}
where \DSLrev{$r\equiv q-1$, $P_{\rm graph}(\mathbf{A})$ is in Eq.~(\ref{pg})}, and Tr denotes the sum over 
\DSLrev{all} $q^N$ spin states.
\DSLrev{With no external field ($H=0$), we have} 
$Z_N(q,0)=e^{NK(1-M_2)} \langle q^C\rangle$, 
where $C=\sum_s n_\mathbf{A}(s)$ is the total number of clusters in a graph \DSLrev{of adjacency matrix $\mathbf{A}$}.  

\DSLrev{If we define the magnetization of the Potts model as $m(q,H) \equiv {1\over N\, r}\sum_{i=1}^N \langle q\delta_{\sigma_i, 1}-1\rangle$ with $\langle \cdots \rangle$ the ensemble average as in Eq.~(\ref{ensave}), it  is evaluated in the limit $q\to 1$ as} 
\begin{equation}
m(1,H)
=\lim_{q\to1}{1\over rN} {\partial \over \partial H} \ln Z_N(q,H)
=\sum_{s\ge1}P(s)(1-e^{-sH}), 
\label{m1H}
\end{equation}
where we have introduced the cluster size distribution 
\DSLrev{$P(s)\equiv \sum_{\mathbf{A}} P_{\rm graph}(\mathbf{A}) {s\over N} n_{\mathbf{A}}(s)$ and it is related to the $n_s$ in Eq.~(\ref{sns}) by $n_s=\langle n_{\mathbf{A}}(s)\rangle = P(s)/s$.}
\DSLrev{If  we take the limit $H\to 0$ satisfying  $s \, H\to 0$ for all cluster size $s$ smaller than the average largest cluster size $\langle S\rangle$ and  $\langle S\rangle \, H\to\infty$, 
the magnetization in Eq.~(\ref{m1H}) is} \DSLfinal{related to $\langle S\rangle$ as}
\begin{equation}
m(1,H\to 0)=   {\langle S\rangle \over N}.
\label{m10}
\end{equation}
\DSLrev{In the same limit,} the susceptibility is obtained as 
\begin{equation}
\chi=\lim_{q\to 1}\lim_{H\to 0}\frac{1}{q}\frac{\partial}{\partial H} m(q,H) = \sum_{s<\langle S \rangle} sP(s),
\label{chi}
\end{equation}
\DSLfinal{which is equal to the average size $\bar{s}$ of  clusters, except for the largest cluster.}

The total number of  clusters $C$ is related to the number of loops $N_{\rm loop}$ by
\DSLfinal{$N_{\rm loop}=L-N+C$.}
\DSLrev{Using  Eq.~(\ref{znqh}) with $H=0$, one can evaluate 
the number of loops per node as}
\begin{eqnarray}
{\langle N_{\rm loop}\rangle \over N}={\langle L\rangle\over N}-1
+{1\over N}{\partial \over \partial q}\left[
\ln Z_N(q,0)\right]_{q=1}.
\label{nloop}
\end{eqnarray}
\\

\subsection{Percolation transition of SF networks}
To evaluate \DSLrev{the magnetization, the susceptibility, and the number of loops in Eqs.~(\ref{m1H}), (\ref{chi}), and (\ref{nloop}), respectively,}  the partition function \DSLrev{$Z_N(q,H)$ should be obtained. To this end}, we associate an $r$-dimensional vector $\vec{S}(\sigma_i)$ of unit length to each spin value $\sigma_i$, where $\vec{S}(1)=(1,0,\ldots,0)$ and $\vec{S}(\sigma_i)$ with $\sigma_i=2,3,
\ldots, q$ point to the remaining $r$ corners of the $r$-dimensional \DSLfinal{hypertetrahedron~\cite{wu}.}
Then the Kronecker delta function in Eq.~(\ref{hm}) is represented as a dot product 
between $\vec{S}$'s,
$\delta_{\sigma_i,\sigma_j}={1\over q}(r\, \vec{S}(\sigma_i)\cdot
\vec{S}(\sigma_j)+1).$
Using \DSLfinal{this,} 
we can rewrite the interaction term in Eq.~(\ref{hm})  as 
$2NK\sum_{i>j}p_i p_j \delta_{\sigma_i,\sigma_j}=
NK\left({1\over q}-M_2\right)+{rNK\over q}  
\left(\sum_i p_i \vec{S}(\sigma_i)\right)^2.$
The perfect square term can be linearized \DSLfinal{by using the Gaussian integral with respect to auxiliary variables,} 
which leads the spin vectors to be decoupled in the partition function. Applying the standard saddle point approximation, we find the partition function represented as~\cite{lee04}
\begin{eqnarray}
{1\over N}\ln Z_N(q,H)=\ln q +K \left({1\over q}-M_2\right)-r F(y,H),\nonumber \\
\label{logZ}
\end{eqnarray}
with
\begin{eqnarray}
F(y,H)={q\over 4K}y^2 - {1\over rN}\sum_{i=1}^N \ln \zeta(H+Np_iy,q),
\label{fyh0}
\end{eqnarray}
where
\begin{eqnarray}
\zeta(h,q)={1\over q}\sum_{\sigma=1}^q e^{rhS_1(\sigma)} = 
{e^{rh}+r e^{-h}\over 1+r},
\label{zetah}
\end{eqnarray}
and $y$ is \DSLrev{determined by the condition that $F(y,H)$ is minimum at $y$ or $(\partial/\partial y) F(y,H)=0$}. 
In the $r\to 0$ limit, \DSLrev{$F(y,H)$ in Eq.~(\ref{fyh0}) is represented} \DSLfinal{as~\cite{lee04}}
\begin{eqnarray}
F(y,H)={1\over 4K}y^2-{1\over N}\sum_{i=1}^N (e^{-h_i}-1+h_i), 
\label{fyh02}
\end{eqnarray}
where $h_i=H+Np_iy$ and $y$ is the solution of 
\begin{eqnarray}
{y\over 2K}=\sum_{i=1}^N p_i (1-e^{-h_i}). 
\label{sp3}
\end{eqnarray}
\DSLrev{Using Eqs.~(\ref{logZ}) and ~(\ref{fyh02}) in} 
\DSLfinal{Eq.~(\ref{m1H}),} 
we obtain  the magnetization 
\begin{eqnarray}
m(1,H)={1\over N}\sum_{i=1}^N (1-e^{-h_i}).
\label{mag2}
\end{eqnarray}
\DSLrev{The susceptibility in Eq.~(\ref{chi}) is evaluated as}
\begin{eqnarray}
\chi(1,H)&=&{1\over N}\sum_{i=1}^N e^{-h_i} 
(1+Np_i {dy \over dH}), \nonumber \\
&=&
{1\over N}\sum_{i=1}^N e^{-h_i} + 
{\left(\sum_{i=1}^N p_i e^{-h_i}\right)^2 \over 
	(2K)^{-1}-\sum_{i=1}^N Np_i^2 e^{-h_i}}.
\label{chi2}
\end{eqnarray}

With $H\to 0$, a non-\DSLrev{zero} solution of Eq.~(\ref{sp3}) 
\DSLrev{exists} 
if $(2K)^{-1}<N\sum_{i=1}^N p_i^2$,  which gives the 
\DSLfinal{percolation threshold $K_c = \left(2N\sum_i p_i^2\right)^{-1}$ represented as}
\begin{widetext}
\begin{equation}
K_c
\approx\left\{
\begin{tabular}{ll}$\frac{1-2\mu}{2(1-\mu)^2}$ &\quad $0<\mu<1/2$ \quad ($3 < \lambda < \infty$),\\
$\frac{1}{2(1-\mu)^2\zeta(2\mu)}N^{-(2\mu-1)}$ &\quad $1/2<\mu<1$ \quad ($2 < \lambda <3$). \\
\end{tabular}\right.
\label{kc0}
\end{equation}
\end{widetext}

Also, the number of loops is  
\begin{equation}
{\langle N_{\rm loop}\rangle\over N}= {\langle L \rangle \over N }
-K -F(y,0)|_{q=1},
\label{ell1}
\end{equation}
where $y$ is given by Eq.~(\ref{sp3}) with $H=0$.

\DSLrev{Setting $H= -\ln z$, one finds that 
the magnetization $m(1,H)$ in Eq.~(\ref{m1H}) is related to the generating function $\mathcal{P}(z)$ of the  cluster size distribution $P(s)$ as $\mathcal{P}(z) = \sum_s P(s) z^s = 1- m(1,H=-\ln z)$.}

\subsection{Critical behaviors}

We characterize the critical behaviors depending on the degree exponent $\lambda$. The range of $\mu$ ($\lambda$) is divided into the three regions, (I) $0 \le \mu <1/3$ ($4 < \lambda \le \infty$), (II) $1/3<\mu<1/2$ ($3 < \lambda < 4$), and (III) $1/2<\mu<1$ ($2 < \lambda < 3$). \DSLrev{In these regions, Eq.~(\ref{sp3}) is expanded differently in powers of $y$, giving different critical behaviors and cluster size distributions~ \cite{lee04} as summarized below.}

i) The order parameter: When $K>K_c$ for (I) and (II) or $K>0$ for (III),
\begin{equation}
m \sim \left\{\begin{array}{lll}
\Delta & \qquad ({\rm I})  \\[2.5mm]
\Delta^{\mu\over 1-2\mu} & \qquad ({\rm II}) \\[2.5mm]
K^{\mu\over 2\mu-1} & \qquad ({\rm III}),
\end{array}\right.
\label{ysol}
\end{equation}
where $\Delta\equiv K/K_c-1$ for (I) and (II). 

ii) The susceptibility or the mean cluster size:
\begin{eqnarray}
\DSLfinal{\chi= } \bar{s}\sim  
\left\{\begin{array}{ll}
{\Delta^{-1}}&(\Delta<0)\\
{\Delta^{-1}}&(\Delta>0) 
\end{array}
\right. \quad {\rm \,\, (I) \,\, and \,\, (II)}.
\label{meancluster}
\end{eqnarray}
For (III), the  mean cluster size does not diverge at any value of $K$.

iii) The number of loops: 
When $K>K_c$ for (I) and (II) or $K>0$ for (III), 
\begin{eqnarray}
\DSLfinal{\frac{\langle N_{\rm loop}\rangle}{N}}\simeq\left\{
\begin{array}{ll}
\Delta^3 
& \qquad ({\rm I})\\
\Delta^{1\over 1-2\mu}& \qquad ({\rm II})
\\
K^{1\over 2\mu-1} & \qquad
({\rm III}).
\end{array}
\right.
\label{ell3}
\end{eqnarray}

iv) The \DSLrev{cluster size} distribution $P(s)$ of a node belonging to a cluster of size $s$: 
The distribution $P(s)$ near $K_c$ takes the form
\begin{equation} 
P(s)\sim s^{1-\tau}e^{-s/s_c} 
\end{equation}
with $s_c\sim(K-K_c)^{-1/\sigma}$. The critical exponents
$\tau$ and $\sigma$ are evaluated as shown in Table~\ref{tab:tau}.
\begin{table}
\caption{The critical exponents $\tau$ and $\sigma$ describing the cluster size distribution.}
\centering
\begin{minipage}{1.0\linewidth}
\begin{tabular}{cccc}
\hline
~~~~~~~~~&~~~~~~~~&~~~~~~~~$\tau-1$~~~~~~ & ~~~~~~ $\sigma$~~~~~~~~ \\
\hline
(I) & $0 \le \mu < 1/3$ & $\frac{3}{2}$ & $\frac{1}{2}$ \\[2.0mm]
(II) & $1/3<\mu < 1/2$ & $\frac{1}{1-\mu}$ & $\frac{1-2\mu}{1-\mu}$ \\[2.0mm]
(III) & $1/2<\mu <1$ & $\frac{1}{\mu}$ & $\frac{2\mu-1}{1-\mu}$ \\
\hline
\end{tabular}
\label{tab:tau}
\end{minipage}
\end{table}

v) The largest cluster size $\langle S\rangle$ in finite systems at $K_c(N)$: 
\begin{equation}
\langle S\rangle\sim\left\{\begin{tabular}{ll}
	$N^{1/(\tau-1)}$ & \quad $0 < \mu <1/2$, \\
	$K_c(N)N^{1/(\tau-1)}\sim N^{1-\mu}$ & \quad $1/2 < \mu < 1$. \\
\end{tabular}
\right.
\end{equation}

\section{Explosive percolation}\label{sec:ep}
 
Percolation transitions have been considered to be continuous. Recently, interest in whether discontinuous percolation transition (DPT) exists or not has been triggered and boosted by the explosive percolation (EP) model~\cite{ep}. The EP model, motivated by a mathematical invention, is a variant of the ER model by adopting the so-called Achlioptas process. In this model, a system begins with $N$ isolated nodes. For each edge attachment, two pairs of nodes that are not yet connected are chosen randomly and one of those pairs is taken and connected by some rule. In the rule, the chosen pair produces a smaller connected cluster than the other pair produces. This selection rule can be extended to arbitrary $m$ pairs of node  candidates~\cite{oliveira, tricritical}. Among those $m$ pairs of node candidates, one pair of nodes, which produces the smallest cluster is taken and connected. We refer to this rule as the $m$-optional Achlioptas process. The original EP model is the case $m=2$. The Achlioptas process suppresses the growth of large clusters so that the percolation threshold is delayed and the medium-size clusters become abundant in the system before the threshold. As a result, once a percolation threshold is passed, the size of the largest cluster is drastically increased by coalescence of medium-size clusters~\cite{friedman, hooyberghs, cho_scirep}. Thus, percolation transition of the EP model was regarded as a discontinuous transition when it was first introduced. The authors of Ref.~\cite{ep} provided a simple argument for their claim that the EP model exhibits a discontinuous percolation transition. The EP model was first introduced on a basis of the ER network and was extended to the square lattice~\cite{ziff, ziff_lattice} and scale-free networks~\cite{cho_ys, filippo}. Results obtained from different embedded spaces were similar to that from the ER network. As many variants of the EP model generated abrupt percolation transitions~\cite{boccaletti_review, cho_supp, gaussian, bfw}, it was required to clarify the transition type of the EP model. 

The authors of~\cite{mendes} introduced a modified EP model in which the rate equation for the evolution of cluster sizes can be constructed. Even though this approach could not provide an analytic solution determining the type of EP transition, it could provide a refined numerical value of the critical exponent of order parameter. They obtained the nonzero value $\beta\approx 0.05$ for the modified EP model. Thus, they claimed that the EP transition is actually continuous. However, because the numerical value $\beta \approx 0.05$ is too small, more careful analysis based on another type of EP model was needed. At that stage, two mathematicians argued that the number of clusters that participate in the emergence of a macroscopic-scale giant cluster should be extensive to the system size $N$ for the EP model with local rule (finite $m$)~\cite{riordan}. Thus, they supported the claim that the EP model is actually continuous~\cite{grassberger_cont, hklee_cont}. Moreover, they presented a strong argument that any percolation model with local rule does not guarantee a discontinuous transition. 

The percolation transition in the ER model follows the mean-field solution of ordinary percolation. 
From this perspective, it would be interesting to consider how the EP transition in Euclidean space is related to that on a random graph. Motivated by this, we introduced the so-called spanning-cluster-avoiding (SCA) model~\cite{cho_science}. In this model, the target pattern in the Achlioptas process is taken as a spanning cluster instead of giant cluster, following the convention of percolation in Euclidean space. Specifically, we consider a bond percolation problem on $d$-dimensional Euclidean lattice. A bond that creates a spanning cluster when it is occupied is called a bridge bond. At each time step, $m$ unoccupied bond candidates are chosen randomly; among them, if there is at least one non-bridge bond, we occupy it definitely. In this case, if there are multiple non-bridge bonds among the candidates, we occupy one of them randomly. In the early time steps, occupied bonds are rare, so the density of bridge bonds is zero. With increasing time step, the density of bridge bonds is increased, and the probability for the emergence of a spanning cluster is increased. The order parameter is the fraction of sites that belong to the spanning cluster. Using the scaling formula for the density of bridge bonds~\cite{bridge_scaling}, the percolation threshold of the SCA model could be analytically calculated for any $m$ potential bonds in the Achlioptas process. This analytic result leads to the following conclusion: the EP transition is continuous for $m < m_c$ and discontinuous for $m > m_c$, where $m_c\sim \ln N$ if the spatial dimension is less than the upper critical dimension $d_c=6$, and the EP transition is always continuous otherwise. Subsequently, it was concluded that the transition of the original EP model~\cite{ep} is continuous as a mean-field solution of the SCA model. 

\section{Hybrid percolation transition}
\label{sec:hybrid}

A hybrid phase transition is a type of phase transition exhibiting properties of both second-order (critical phenomena) and first-order (jump of the order parameter) phase transitions at the same transition point. In spin systems, such a type of phase transition occurs at the so-called critical endpoint in the systems with competing interactions such as the Ashkin-Teller model on scale-free networks~\cite{mukamel, at}. Recently, similar hybrid percolation transitions (HPTs) have been obtained, for instance, restricted percolation model~\cite{half,r_ER_hybrid,kchoi_hybrid}, $k$-core percolation~\cite{kcore1,kcore2,kcore3, kcore_prx} and the cascade failure (CF) model on interdependent networks \cite{dodds,janssen,grassberger_epi}. For such models, the order parameter $m(t)$ behaves as    
\begin{equation}
m(t)=\left\{
\begin{array}{lr}
0 & ~{\rm for}~~  t < t_c, \\
m_0+r(t-t_c)^{\beta_m} & ~{\rm for}~~ t \ge t_c, 
\end{array}
\right.
\label{eq:hpt_order}
\end{equation}
where $m_0$ and $r$ are constants, $\beta_m$ is the critical exponent of the order parameter, and $t$ is a control parameter such as the number of edges per node and the mean degree of a given network. In such cases, the HPT occurs at $t_c$ as edges are added (or deleted) one by one following a given rule from a certain point far below (above) the percolation threshold. Such a transition is called the HPT in cluster merging (pruning) processes. 

\subsection{HPTs in cluster merging processes}

As we described, a discontinuous percolation transition in the cluster merging process can occur when a global rule is considered. For example, for the SCA model, one has to check whether each bond candidate can make a spanning cluster if it is attached~\cite{cho_science}. Another example is the model in which a discontinuous percolation transition is generated by controlling only the largest cluster~\cite{largest}. That is, one needs global information to generate a discontinuous percolation transition. However, while the order parameter is increased rapidly in such discontinuous percolation transitions, critical behavior hardly appears. Thus, the question of whether an HPT can occur in cluster merging processes was raised. Recently, the authors of Ref.~\cite{r_ER_hybrid} slightly modified an existing model~\cite{half} and successfully generated a hybrid percolation transition.   

The model is defined as follows: At the beginning, the system consists of $N$ isolated nodes. At each time step, we first rank the clusters by ascending order of cluster size. If multiple clusters of the same size exist, they are randomly sorted. The restricted set of clusters $R(t)$ at time $t$ is defined as the subset consisting of a certain number of smallest clusters (say $k$ clusters) and is denoted as $R(t) \equiv \{c_1, c_2, \cdots, c_k\}$. Further, $k$ is determined as the value satisfying the inequalities $N_{k-1}(t) < \lfloor gN \rfloor \le N_k(t)$ for a given model parameter $g\in (0,1]$. $N_k(t) \equiv \sum_{\ell=1}^k s_{\ell}(t)$, where $s_{\ell}(t)$ is the number of nodes in the cluster $c_{\ell}$. We note that the number of nodes in $R(t)$ varies with the time step $t$. Here the time step $t$ is defined as the number of edges added to the system per node. 
This model is called a restricted ER model, because when $g=1$, the model is reduced to percolation in the ordinary ER model. We remark that this restricted ER model is a slightly modified version of the original model~\cite{half} in which the number of nodes in the set $R$ is fixed as $\lfloor gN \rfloor$. Thus, some nodes in a cluster on the boundary between the two sets $R$ and $R^{(c)}$ belong to the set $R$, and the others in the same cluster belong to the set $R^{(c)}$. However, for the modified model, all the nodes in the cluster are counted as elements of the set $R$. This modification enables one to solve analytically the phase transition for $t > t_c$ without changing any critical properties.

This restricted ER model exhibits an HPT at a transition point $t_c$. The order parameter $m(t)$, that is, the fraction of nodes belonging to the giant cluster, increases rapidly from zero at $t_c^{-}$ to a finite value $m_0$ at $t_c$. The interval $\Delta t=t_c-t_c^- \sim o(N)/N$. Thus, in the thermodynamic limit, this interval reduces to zero, and the order parameter is regarded as jumping discontinuously at $t_c$. For $t> t_c$, $m(t)$ increases gradually following formula~(\ref{eq:hpt_order}). Moreover, the size distribution of finite clusters, $n_s(t)$, exhibits power-law decay at $t_c$ with the exponent $\tau(g)$ in the range $2 < \tau(g) \le 2.5$. Thus, the critical exponents of the HPT vary continuously depending on the control parameter $g$. Such critical behaviors of the HPT in the cluster merging process have been observed for the first time.   

\begin{figure}
\centering
\includegraphics[width=.96\linewidth]{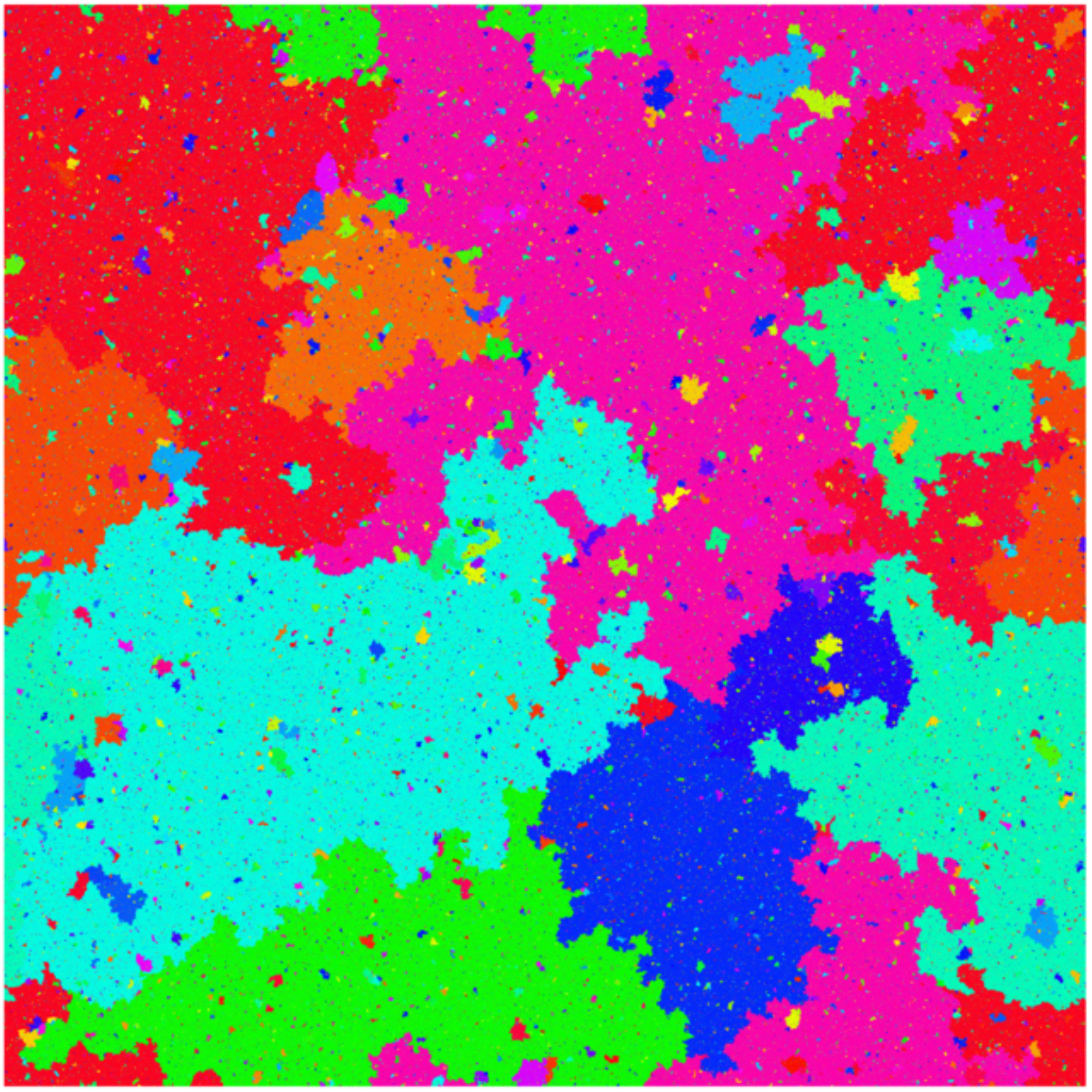}
\caption{A snapshot of the restricted percolation model with $g = 0.5$ on two dimensional square lattice at the time step $t$ just before the order parameter jumps from zero to a finite value.
}
\label{fig2}
\end{figure}

Recently, we investigated the critical behaviors of the HPT arising in the restricted percolation model in two dimensions using finite-size scaling theory~\cite{kchoi_hybrid}. We obtained the  correlation exponent $\nu_g$ which determines the transition point of finite systems using the scaling relation. We found that the exponent satisfies the hyperscaling relation with other critical exponents. Moreover, we found that the size distribution of finite clusters follows a power law at the transition point with exponential cutoff due to finite size effect. Another correlation length exponent $\nu_s$ is necessary to characterize the finite-size scaling behavior of the size distribution of finite clusters. We noticed that the two correlation length exponents $\nu_g$ and $\nu_s$ are different. In fact, they are the same in the second-order transition. We examined the shapes of giant and finite clusters, which are almost compact with the fractal dimension $D_g=D_s=2=d$ (Fig.~\ref{fig2}). This seems to be related to the property $\beta=0$ of the first-order transition. On the other hand, in equilibrium spin models, the critical behavior can be understood using the renormalization group (RG) transformation of the singular part of the free energy function $f$. That is, $f(t,h)=\ell^{-d}f(t^{\prime}, h^{\prime})$, where $t$ and $h$ are reduced temperature and external field, respectively and $t^{\prime}=\ell^{y_t} t$ and $h^{\prime}=\ell^{y_h} h$, where $y_t=1/\nu$, $y_h=D$, and $\ell$ is the scale factor of coarse-grained length. $D$ is the fractal dimension satisfying $D=d-\beta/\nu$. On the other hand, for a discontinuous transition, $y_t=y_h=d$, because of $\beta=0$. In short, for the HPT, there exist two $\nu$ exponents $\nu_g$ and $\nu_s$, and $D=d$. While those two correlation length exponents satisfy their own hyperscaling relations with other critical exponents, it seems that we need a single theoretical framework for the HPT within which those critical exponents are unified.  

\subsection{HPTs induced by cascading dynamics}

The cascading failure (CF) model on interdependent networks~\cite{buldyrev,son_grassberger,baxter,bashan,bianconi,zhou,makse,boccaletti,kivela} and $k$-core percolation~\cite{kcore1, kcore_jamming, kcore2, kcore_mendes_2, kcore3} are the paradigmatic models of the HPTs in pruning processes. In this CF model the process is controlled by the mean degree $t$ of the networks. When a node on one layer fails and is deleted, it leads to another failure of the counterpart node in the other layer of the network. Subsequently, links connected to the deleted nodes are also deleted from the networks. This process continues back and forth, always eliminating the possibly separated finite clusters until a giant mutually connected component (GMCC) remains or the giant component gets entirely destroyed as a result of the cascades. As nodes are deleted in such a way, the behavior is similar to that at a second-order phase transition until the transition point $t_c$ is reached from above. Beyond that, as $t$ is further decreased infinitesimally, the percolation order parameter drops suddenly to zero indicating a first-order phase transition. Thus a HPT occurs at $t = t_c$ as shown in Fig.~1. This transition may be regarded as a transition to an absorbing state. 

The $k$-core of a network is the connected cluster of nodes in which each node has at least $k$ neighbors in the cluster. If a node is removed from the $k$-core, its neighbors with exactly $k$ neighbors in the $k$-core must be deleted from the $k$-core. Thus, similar to the CF model, removal of a node leads to another removal of nodes successively, and the size of the $k$-core is reduced.

For the phase transitions in these models, the mean degree $t$ of the network is the control parameter and the size $m$ of the giant cluster (GMCC or $k$-core) relative to the system size is the order parameter. Here we consider only ER networks. The process starts from a network with a sufficiently large mean degree where a giant cluster exists. The relative size of the giant cluster decreases as the edges are deleted one by one, and it suddenly collapses from a finite value to zero at a certain critical mean degree $t_c$. Thus the phase transition is discontinuous. However, the process near $t_c$ also shows singularities that are usually found in continuous phase transitions. For instance, the leading term of $m(t)$ as a function of the mean degree $t$ near $t_c$ is described as Eq.~(\ref{eq:hpt_order}) with a fractional power $\beta_m = 1/2$.

The deletion of an edge or a node causes an avalanche of the separations of nodes from the giant cluster. This can be viewed as the response on the perturbation in complex systems. The mean size of the avalanche is finite if $t > t_c$, i.e., the number of separated nodes are finite even in the thermodynamic limit and the relative size of the giant cluster also remains finite after the avalanche. However, as $t$ approaches $t_c$, the mean avalanche size diverges and the giant cluster can vanish entirely by a single avalanche. 
The size and duration time of the avalanche and the size of the giant cluster are the main quantity of interest. The avalanche size distribution follows a power law and the size distribution of the giant cluster is bimodal near $t_c$. The duration time of the infinite avalanche exhibits a scaling on the system size as $t_d \sim O(N^{1/3})$, which is universal across diverse models. Their numerical values and relations encode the essence of the HPTs. Unlike the ordinary percolation, the cluster size distribution is trivial in these models. Each node outside the giant cluster is isolated, and thus the cluster size distribution does not play an important role.

To describe the HPT of the CF model and $k$-core on the interdependent networks, we need two sets of critical exponents. The set $\{\beta_m, \gamma_m, {\bar \nu}_m \}$ is to describe the behavior of the order parameter, and the other set, $\{\tau_a, \sigma_a, \gamma_a, {\bar \nu}_a \}$, is to describe the behavior of the avalanche size distribution. Here the subscripts $m$ and $a$ refer to the order parameter and avalanche, respectively.
The exponent $\beta_m$ describes the growth of the order parameter near $t_c$ and defined with Eq.~(\ref{eq:hpt_order}). The finite size scaling of the order parameter $m - m_0 \sim N^{-\beta_m / {\bar \nu}_m}$ defines ${\bar \nu}_m$. The exponent $\gamma_m$ is for the fluctuation of the order parameter: $N (\left< m^2 \right> - \left< m \right>^2) \sim (t - t_c)^{-\gamma_m}$.
The exponent $\tau_a$ and $\sigma_a$ are defined by the finite avalanche size distribution $p_a(s) \sim s^{-\tau_a} \mathcal{G}(s/s_*)$, where $s$ is the avalanche size and $\mathcal{G}$ expresses the cutoff at $s_{*} \sim (t - t_c)^{-1/\sigma_a}$. The mean size of the finite avalanche $\left< s \right> \sim (t - t_c)^{-\gamma_a}$ and its finite size scaling $\left< s \right> \sim N^{-\gamma_a / \bar{\nu}_a}$ define the exponents $\gamma_a$ and $\bar{\nu}_a$. The numerical values of the exponents are reported in \cite{mcc_lee} for the CF model and \cite{kcore_lee} for the $k$-core percolation.

One could expect that the power-law behavior of the avalanche size distribution in these models may replace the role of the power-law behavior of the cluster size distribution in the ordinary percolation. In that case the two exponents $\gamma_m$ and $\gamma_a$ must have the same value. However, it turns out that they are different in both models. Instead, the two sets of exponents are related by a scaling relation $1-\beta_m=\gamma_a$. This is because the mean size of the finite avalanche is essentially the mean decrement of the size of the giant cluster by an infinitesimal decrement of the mean degree which is implemented as a deletion of an edge~\cite{mcc_lee,kcore_lee}.
The exponents $\bar{\nu}_m$ and $\bar{\nu}_a$ represent the behavior of the size scales of the models similarly to the correlation length exponent of the models in Euclidean spaces.
The two exponents in the CF model have different values while they have the same value in the $k$-core percolation. This indicates that the CF model has two different size scales diverging at $t_c$ but it is not a general property of the HPT. 

An important perspective to understand the ordinary percolation is to study it in terms of the branching process. This idea is also important for the HPTs. The hybrid nature of the transition is represented as a crossover from a critical branching process to a supercritical branching process in a single infinite avalanche~\cite{universal_mech} while a giant cluster of the ordinary percolation near $t_c$ is described as a simple critical branching process. The mechanism underlying the crossover is universal over various models~\cite{universal_mech} including not only the pruning processes of the CF model and the $k$-core percolation but also epidemic models such as SWIR model~\cite{janssen,grassberger_epi,swir_choi_1,swir_choi_2} and the threshold model~\cite{dodds}.
Removal of a node from the giant cluster triggers an avalanche which is basically a branching process.
The branching process near $t_c$ forms a critical process in its early stage. Nodes are separated from the giant cluster by the critical process but some nodes remaining in the giant cluster become vulnerable during the process. As time goes on, the branching process forms large loops owing to the finite size of the network and the accumulated vulnerable nodes are eventually separated by the branching process.
This later stage of the process is supercritical. The mean branching ratio is larger than unity and the avalanche size grows exponentially during a short time. For the CF model and $k$-core percolation, the vulnerability of a node is measured in terms of the number of neighbors in the giant cluster.
The time to the crossover scales as $N^{1/3}$ in ER networks. The scaling is independent of the details of the dynamic rules~\cite{universal_mech} and closely related to the fact that the giant cluster of ER network near $t_c$ is locally tree-like and its linear size is order of $N^{1/3}$~\cite{krapivsky}.

\section{Mutual percolation and related percolation problems in multiplex networks}
\label{sec:mcc}

The CF model on interdependent networks \cite{buldyrev} invokes explicitly the dynamical steps of cascades. An equivalent formulation based purely on graph-topological notions \cite{Son2012}, commonly known in the name of ``mutual percolation,'' has also been studied quite extensively on multiplex networks \cite{Lee2015}. In this formulation, the key graph-theoretic notion is the mutual connectivity. 
Given a multiplex network of more than one layers where nodes are linked via more than one type of links, a pair of nodes is said to be mutually connected, if there exists a connected path between the two nodes in each and every layer.  The set of nodes for which every pair is mutually connected defines the mutually connected component or mutual component, for short. The extensive mutual component, if exists, is called the giant mutually connected component (GMCC) or giant mutual component. The mutual percolation problem deals with the existence of GMCC in multiplex networks and the associated phase transition and critical phenomena. 

The mutual percolation on multiplex random networks is amenable to analytic treatment based on generalization of the mean-field-type argument developed for single-layer percolation problems \cite{Son2012}. It has spawned many studies of variations of the model; to name just a notable few, the partially-mutual percolation \cite{Son2012,Parshani2010}, the antagonistic percolation \cite{Zhao2013}, the redundant interdependency model \cite{Radicchi2017}, not to mention the studies of mutual percolation on various network settings \cite{Min2014a,makse,Hu2013,bianconi,Min2015}. As seen from the results of CF model, a key feature of the mutual percolation and its variants is the generic emergence of discontinuous, often hybrid, percolation transition for the emergence of GMCC (or its generalized counterpart). Despite numerous effort, the physics behind it is still not fully understood. Specifically, how and why discontinuity and critical scaling co-appear, what different critical fluctuations responsible for them are, and how the two are intrinsically related, if ever, are not fully clear yet. 

The presence of discontinuity suggests possibility of bistability of percolating cluster. A few models have been proposed that display bistable percolating cluster based on the mutual percolation \cite{Zhao2013,Min2014b,Baxter2014}. 
In Ref.~\cite{Min2014b}, the notion of multiplex viability was introduced and two different percolation processes establishing the viability were proposed to display the bistability, which could be interpreted as the hysteresis arising from the forward (activating) and backward (deactivating) processes. The mutual percolation was identified, at least formally for the case on random networks, as a particular limit of the deactivating percolation process of the multiplex viability model \cite{Min2014b}, opening the possibility for studying the critical properties of the mutual percolation via the multiplex viability model. 

Mutual percolation is a simplest structural phase transition model for multiplex networks with cooperatively- or conjunctively-coupled layers, and as such serves as the key problem towards further understanding of multiplex complex system dynamics with cooperative layers \cite{Brummitt2012,Lee2014}. Classical percolation problems can also be studied on multiplex networks \cite{Leicht2009,Lee2012,Hackett2016}, which are relevant for problems with complimentary or disjunctively coupled layers \cite{DeDomenico2014,Min2016}.

\section{Summary and Discussion} 
\label{sec:summary}

Since the two pioneering models, the WS model and the BA model, were published in late last century, the subject of complex networks impacted diverse issues in social and scientific area~\cite{net_sci_book} in a short time period. The citation number of each paper exceeds more than 30,000 times during the past 20 years as of April 2018, which may reflect the scientific impacts of those two papers.    
    
Among diverse issues in network science, in this review article we focused on the percolation transition of complex networks. Using the static model, we constructed SF random graphs in which edges are attached between two nodes with the probability proportional to the product of two vertex weights. This model enables us to calculate structural properties of percolating clusters using the Potts model formalism. We presented the method how to derive thermodynamic quantities including the critical exponents of percolation transition of SF networks. 

Percolation transition is known as one of the most robust continuous transitions. We introduced several models which exhibit a discontinuous or hybrid percolation transition. Those models include global dynamic rule, which is indeed the necessary condition to generate such types of percolation transitions. 

Cascading dynamics is another important factor to generate a discontinuous or hybrid phase transition. We introduced several models which exhibit such types of percolation transitions induced by cascading dynamics. We showed some universal behaviors in those models. 

In this review article, we limit our interest to the static case, in which the number of nodes in a system is fixed throughout the dynamics. In the real world, however, the number of nodes in the system can change with time. For instance, the coauthorship network~\cite{coauthorship} and the protein interaction network contain such a factor. In these growing networks, percolation transition is known as an infinite-order transition~\cite{callaway,pin1,pin2,pin3}, belonging to the so-called Berezinskii-Kosteritz-Thouless (BKT) universality class~\cite{rmp_2017,kt,rmp_quantum}. Recent works showed that the BKT transition is vulnerable when the growth of large clusters is suppressed and abnormal phase transitions occur~\cite{oh1,oh2}.  

Finally, since the two papers~\cite{ws,ba99} on complex networks were published in  late last century, a huge number of papers regarding percolation transition have been published. Thus, it is almost impossible to trace them all comprehensively. Here we have reviewed papers based on our publications ranging from the ordinary percolation transition on SF networks to abrupt percolation transitions such as the EP models and hybrid percolation models from our viewpoints. Therefore, many important papers are missing in the main text and the list of references. More detailed reviews of the explosive percolation and the hybrid percolation, and open challenges can be found in Refs~\cite{souza_nphy,nuno,jstat}.

\begin{acknowledgments}
This work was supported by the National Research Foundation of Korea by Grant No. 2014R1A3A2069005 (BK), No. 2016R1A2B4013204 (DSL), No. 2017R1A2B2003121 (KIG), 2017R1C1B1004292 (YSC), and 2017R1A6A3A11033971 (DL).
BK thanks K. Choi, H. J. Herrmann, J. Kert\'esz, D. Kim, J. S. Lee, S. M. Oh and S.-W. Son for their collaborations on the subject related to this review article. 
\end{acknowledgments}


\begin{thebibliography}{99}
\bibitem{complex1} K. Ziemelis and L. Allen, Nature (London) {\bf 410,} 241 (2001).
\bibitem{complex2} R. Gallagher and T. Appenzeller, Science {\bf 284,} 87 (1999).
\bibitem{parisi} G. Parisi, Physica A (Amsterdam) {\bf 263,} 557 (1999).
\bibitem{barkhausen} J. P. Sethna, K. A. Dahmen, and C. R. Myers, Nature (London) {\bf 410,} 242 (2001).
\bibitem{finance} H. E. Stanley et al., Proc. Natl. Acad. Sci. U.S.A. {\bf 99,} 2561 (2002).
\bibitem{buchanan} M. Buchanan, Nature (London) {\bf 419,} 787 (2002).
\bibitem{ws} D. J. Watts, and S. H. Strogatz, Nature (London) {\bf 393}, 440 (1998).
\bibitem{ba99} A.-L. Barab\'{a}si, and R. Albert, Science {\bf 286}, 509 (1999).
\bibitem{review1} R. Albert, and A.-L. Barab\'{a}si, Rev. Mod. Phys. {\bf 74}, 47 (2002).
\bibitem{review2} S. N. Dorogovtsev, and J. F. F. Mendes, Adv. Phys. {\bf 51}, 1079 (2002). 
\bibitem{review3} M. E. J. Newman, SIAM Rev. {\bf 45}, 167 (2003).
\bibitem{er} P. Erd\H{o}s, and A. R\'{e}nyi, Publ. Math. Inst. Hung. Acad. Sci. {\bf 5}, 17 (1960).
\bibitem{www} R. Albert, H. Jeong, A.-L. Barab\'{a}si, Nature {\bf 401}, 130 (1999).
\bibitem{Internet} M. Faloutsos, P. Faloutsos, C. Faloutsos, Comput. Commun. Rev. {\bf 29}, 251 (1999).
\bibitem{diameter_change} J.-H. Kim, K.-I. Goh, B. Kahng, and D. Kim, Phys. Rev. Lett. {\bf 91,} 058701 (2003).
\bibitem{resilience2} R. Albert, H. Jeong, A. L. Barabási, Nature {\bf 406}, 378 (2000).
\bibitem{cohen00} R. Cohen, K. Erez, D. ben-Avraham, and S. Havlin, Phys. Rev. Lett. {\bf 85}, 4626 (2000). 
\bibitem{redner} P. L. Krapivsky, S. Redner, and F. Leyvraz, Phys. Rev. Lett. {\bf 85}, 4629 (2000).
\bibitem{mendes_percolation} S. N. Dorogovtsev, J. F. F. Mendes, and A. N. Samukhin, Phys. Rev. Lett. {\bf 85}, 4633 (2000).
\bibitem{lee04} D.-S. Lee, K.-I. Goh, B. Kahng, and D. Kim, Nucl. Phys. B. {\bf 696}, 351 (2004).
\bibitem{p_report} S. Boccalettia, V. Latora, Y. Moreno, M. Chavez, and D.-U. Hwang, Phys. Rep. {\bf 424}, 175 (2006).
\bibitem{ultra_small} R. Cohen, and S. Havlin, Phys. Rev. Lett. {\bf 90}, 058701 (2003).
\bibitem{molloy1} M. Molloy, and B. Reed, Random Struct. Algorithms {\bf 6}, 161 (1995).
\bibitem{molloy2} M. Molloy, and B. Reed, Combinatorics, Probab. Comput. {\bf 7}, 295 (1998).
\bibitem{static_model} K.-I. Goh, B. Kahng, and D. Kim, Phys. Rev. Lett. {\bf 87}, 278701 (2001).
\bibitem{chung} F. Chung, and L. Lu, Annals Combinatorics {\bf 6}, 125 (2002).
\bibitem{hammersley_1954} S. R. Broadbent, and J. M. Hammersley, Cambridge Philos. Soc. {\bf 53}, 629 (1957).
\bibitem{flory1} P. J. Flory, J. Am. Chem. Soc. {\bf 63}, 3083 (1941).
\bibitem{flory2} P. J. Flory, J. Am. Chem. Soc. {\bf 63}, 3091 (1941).
\bibitem{flory3} P. J. Flory, J. Am. Chem. Soc. {\bf 63}, 3096 (1941).
\bibitem{disease} J. D. Murray, {\em Mathematical Biology, 3rd edn.} (Springer, Berlin) (2005).
\bibitem{con_insul} D. S. McLachlan, M. Blaszkiewicz, and R. E. Newnham, J. Am. Ceram. Soc. {\bf 73}, 2187 (1990).
\bibitem{schulman_star_perc} L. S. Schulman, and P. E. Seiden, Science {\bf 233}, 425 (1986).
\bibitem{dilute_magnet} L. Bergqvist, O. Eriksson, J. Kudrnovsk\'{y}, V. Drchal, P. Korzhavyi, and I. Turek Phys. Rev. Lett. {\bf 93}, 137202 (2004).
\bibitem{resilience1} R. Albert, H. Jeong, and A. L. Barab\'asi, Nature {\bf 406}, 378 (2000).
\bibitem{resilience3} F. Morone, and H. A. Makse, Nature {\bf 524}, 65 (2015).
\bibitem{opinion1} D. J. Watts, Proc. Natl. Acad. Sci. {\bf 99}, 5766 (2002).
\bibitem{opinion2} J. Shao, S. Havlin, and H. E. Stanley, Phys. Rev. Lett. {\bf 103}, 018701 (2009).
\bibitem{nonvolatile_memory_1} R. Degraeve, G. Groeseneken, R. Bellens, M. Depas, and H. E. Maes {\em Tech. Dig. 1995 Inter. Electron Dev. Meeting} 863 (1995).
\bibitem{nonvolatile_memory_2} R. Degraeve, {\em et al.}, IEEE Trans. Electron Dev. {\bf 51}, 1392 (2004).
\bibitem{stauffer} D. Stauffer, and A. Aharony, {\em Introduction to Percolation Theory, 2nd edn.} (Taylor, and Francis, London) (1994).
\bibitem{fisher1} M. E. Fisher, and J. W. Essam, J. Math. Phys. {\bf 2}, 609 (1961).
\bibitem{fisher2} M. E. Fisher, J. Math. Phys. {\bf 2}, 620 (1961).
\bibitem{kasteleyn} P. W. Kasteleyn, and C. M. Fortuin J. Phys. Soc. Jpn. (Suppl.) {\bf 26}, 11 (1969).
\bibitem{berg02} J. Berg, and M. L\"{a}ssig, Phys. Rev. Lett. {\bf 89}, 228701 (2002).
\bibitem{manna03} M. Baiesi, and S. S. Manna, Phys. Rev. E {\bf 68,} 047103 (2003).
\bibitem{burda} Z. Burda, J. D. Correia, and A. Krzywicki, Phys. Rev. E {\bf 64}, 046118 (2001). 
\bibitem{doro03} S. N. Dorogovtsev, J. F. F. Mendes, and A. N. Samukhin, Nucl. Phys. B {\bf 666,} 396 (2003).
\bibitem{farkas} I. Farkas, I. Derenyi, G. Palla, and T. Viscek, Lecture notes in Physics: {\it Networks: structure, dynamics, and function}, edited by E. Ben-Naim, H. Frauenfelder, and Z. Toroczkai (Springer, 2004).
\bibitem{newman01} M. E. J. Newman, S. H. Strogatz, and D. J. Watts, Phys. Rev. E {\bf 64}, 026118 (2001).
\bibitem{caldarelli02} G. Caldarelli, A. Capocci, P. De Los Rios, and M. A. Mu\~noz, Phys. Rev. Lett. {\bf 89}, 258702 (2002).
\bibitem{soderberg02} B. S\"{o}derberg, Phys. Rev. E {\bf 66}, 066121 (2002).
\bibitem{aiello02} W. Aiello, F. Chung, and L. Lu, Exp. Math. {\bf 10}, 53 (2001).
\bibitem{wu} F. Y. Wu, Rev. Mod. Phys. {\bf 54}, 235 (1982).
\bibitem{cohen02} R. Cohen, D. ben-Avraham, and S. Havlin, Phys. Rev. E {\bf 66}, 036113 (2002).
\bibitem{ep} D. Achlioptas, R.M. D'Souza, J. Spencer, Science {\bf 323}, 1453 (2009). 
\bibitem{oliveira} J. S. Andrade, H. J. Herrmann, A. A. Moreira, and C. L. N. Oliveira, Phys. Rev. E {\bf 83}, 031133 (2011).
\bibitem{tricritical} N. A. M. Ara\'ujo, J. S. Andrade, R. M. Ziff, and H. J. Herrmann, Phys. Rev. Lett. {\bf 106}, 095703 (2011).
\bibitem{friedman} E. J. Friedman, and A. S. Landsberg, Phys. Rev. Lett. {\bf 103}, 255701 (2009).
\bibitem{hooyberghs} H. Hooyberghs, and B. V. Schaeybroeck, Phys. Rev. E {\bf 83}, 032101 (2011).
\bibitem{cho_scirep} Y. S. Cho, and B. Kahng, Sci. Rep. {\bf 5}, 11905 (2015).
\bibitem{ziff} R. M. Ziff, Phys. Rev. Lett. {\bf 103}, 045701 (2009).
\bibitem{ziff_lattice} R. M. Ziff, Phys. Rev. E {\bf 82}, 051105 (2010).
\bibitem{cho_ys} Y. S. Cho, J. S. Kim, J. Park, B. Kahng, and D. Kim Phys. Rev. Lett. {\bf 103}, 135702 (2009).
\bibitem{filippo} R. Filippo, and F. Santo, Phys. Rev. Lett. {\bf 103}, 168701 (2009).
\bibitem{boccaletti_review} S. Boccaletti {\em et al.} Phys. Rep. {\bf 424} (2006). 
\bibitem{cho_supp} Y. S. Cho, and B. Kahng, Phys. Rev. Lett. {\bf 107}, 275703 (2011).
\bibitem{gaussian} K. J. Schrenk, N. A. M. Araujo, and H. J. Herrmann, Phys. Rev. E {\bf 84}, 041136 (2011).
\bibitem{bfw} K. J. Schrenk, A. Felder, and S. Deflorin, Phys. Rev. E {\bf 85}, 031103 (2012).
\bibitem{mendes} R. A. da Costa, S. N. Dorogovtsev, A. V. Goltsev, and J. F. F. Mendes Phys. Rev. Lett. {\bf 105}, 255701 (2010).
\bibitem{riordan} O. Riordan, and L. Warnke Science {\bf 333}, 322 (2011).
\bibitem{grassberger_cont} P. Grassberger, C. Christensen, G. Bizhani, Phys. Rev. Lett. {\bf 106}, 225701 (2011).
\bibitem{hklee_cont} H. K. Lee, B. J. Kim, and H. Park, Phys. Rev. E {\bf 84}, 020101 (2011).
\bibitem{cho_science} Y. S. Cho, S. Hwang, H. J. Herrmann, and B. Kahng Science {\bf 339}, 1185 (2013).
\bibitem{bridge_scaling} K. J. Schrenk, N. A. M. Ara\'{u}jo, J. S. Andrade Jr, and H. J. Herrmann, Sci. Rep. {\bf 2}, 348 (2012).
\bibitem{mukamel} A. Bar, and D. Mukamel, Phys. Rev. Lett. {\bf 112}, 015701 (2014).
\bibitem{at} S. Jang, J. S. Lee, S. Hwang, and B. Kahng, Phys. Rev. E {\bf 92}, 022110 (2015).
\bibitem{half} K. Panagiotou, R. Sph\"{o}el, A. Steger, and H. Thomas, Elec. Notes Discret. Math. {\bf 38}, 699 (2011).
\bibitem{r_ER_hybrid} Y. S. Cho, J. S. Lee, H. J. Herrmann, and B. Kahng, Phys. Rev. Lett. {\bf 116}, 025701 (2016).
\bibitem{kchoi_hybrid} K Choi, D. Lee, Y. S. Cho, J. C. Thiele, H. J. Herrmann, and B. Kahng, Phys. Rev. E {\bf 96}, 042148 (2017).
\bibitem{kcore1} J. Chalupa, P. L. Leath, and G. R. Reich, J. Phys. C {\bf 12}, L31 (1981).
\bibitem{kcore2} S. N. Dorogovtsev, A. V. Goltsev, and J. F. F. Mendes, Phys. Rev. Lett. {\bf 96}, 040601 (2006).
\bibitem{kcore3} A. V. Goltsev, S. N. Dorogovtsev, and J. F. F. Mendes Phys. Rev. E {\bf 73}, 056101 (2006).
\bibitem{kcore_prx} G. J. Baxter, S. N. Dorogovtsev, K. E. Lee, J. F. F. Mendes, and A. V. Goltsev, Phys. Rev. X. {\bf 5}, 031017 (2015).
\bibitem{dodds} P. S. Dodds, and D. J. Watts, Phys. Rev. Lett. {\bf 92}, 218701 (2004).
\bibitem{janssen} H-K Janssen, M. M\"uller, and O. Stenull, Phys. Rev. E {\bf 70}, 026114 (2004).
\bibitem{grassberger_epi} W. Cai, L. Chen, F. Ghanbarnejad, and P. Grassberger, Nat. Phys. {\bf 11}, 936  (2015).
\bibitem{largest} N. A. M. Ara\'ujo, and H. J. Herrmann, Phys. Rev. Lett. {\bf 105}, 035701 (2010).
\bibitem{buldyrev} S. V. Buldyrev, R. Parshani, G. Paul, H. E. Stanley, and S. Havlin, Nature {\bf 464}, 1025 (2010).
\bibitem{son_grassberger} S-W Son, P. Grassberger, and M. Paczuski, Phys. Rev. Lett. {\bf 107}, 195702 (2011).
\bibitem{baxter} G. J. Baxter, S. N. Dorogovtsev, A. V. Goltsev, and J. F. F. Mendes, Phys. Rev. Lett. {\bf 109}, 248701 (2012).
\bibitem{bashan} A. Bashan, Y. Berezin, S. V. Buldyrev, and S. Havlin, Nat. Phys. {\bf 9}, 667  (2013).
\bibitem{bianconi} D. Cellai, E. L\'{o}pez, J. Zhou, J. P. Gleeson, and G. Bianconi, Phys. Rev. E {\bf 88}, 052811 (2013).
\bibitem{zhou} D. Zhou, A. Bashan, R. Cohen, Y. Berezin, N. Shnerb, and S. Havlin, Phys. Rev. E {\bf 90}, 012803 (2014).
\bibitem{makse} S. D. S. Reis, Y. Hu, A. Babino, J. S. S. Andrade Jr, S. Canals, M. Sigman, and H. A. Makse, Nat. Phys. {\bf 10}, 762 (2014).
\bibitem{boccaletti} S. Boccaletti, G. Bianconi, R. Criado, C. I. del Genio, J. G\'omez-Garde\ nes, M. Romance, I. Sendi\~na-Nadal, Z. Wang, and M. Zanin, Phys. Rep. {\bf 544}, 1 (2014).
\bibitem{kivela} M. Kivel\"{a}, A. Arenas, M. Barthelemy, J. P. Gleeson, Y. Moreno, and M. A. Porter, J. Complex Netw. {\bf 2(3)} 203 (2014).
\bibitem{kcore_jamming} J. M. Schwarz, A. J. Liu, and L.Q. Chayes, EPL {\bf 73,} 560 (2006).
\bibitem{kcore_mendes_2} A. V. Goltsev, S. N. Dorogovtsev, and J. F. F. Mendes, Phys. Rev. E {\bf 73}, 056101 (2006).
\bibitem{mcc_lee} D. Lee, S. Choi, M. Stippinger, J. Kert\'esz, and B. Kahng, Phys. Rev. E {\bf 93}, 042109  (2016).
\bibitem{kcore_lee} D. Lee, M. Jo, and B. Kahng, Phys. Rev. E {\bf 94}, 062307 (2016).
\bibitem{universal_mech} D. Lee, W. Choi, J. Kert\'esz, and B. Kahng, Sci. Rep. {\bf 7}, 5723 (2017).
\bibitem{swir_choi_1} W. Choi, D. Lee, and B. Kahng, Phys. Rev. E {\bf 95}, 022304 (2017).
\bibitem{swir_choi_2} W. Choi, D. Lee, and B. Kahng, Phys. Rev. E {\bf 95}, 062115 (2017).
\bibitem{krapivsky} E. Ben-Naim, and P. L. Krapivsky, Phys. Rev. E {\bf 71}, 026129  (2005).
\bibitem{Son2012} S.-W. Son, G. Bizhani, C. Christensen, P. Grassberger, M. Paczuski, EPL {\bf 97}, 16006 (2012).
\bibitem{Lee2015} K.-M. Lee, B. Min, and K.-I. Goh, Eur. Phys. J. B. {\bf 88}, 28 (2015).
\bibitem{Parshani2010} R. Parshani, S.V. Buldyrev, S. Havlin, Phys. Rev. Lett. {\bf 105}, 048701 (2010).
\bibitem{Zhao2013} K. Zhao, G. Bianconi, J. Stat. Mech. {\bf 2013}, P05005 (2013).
\bibitem{Radicchi2017} F. Radicchi, and G. Bianconi, Phys. Rev. X {\bf 7}, 011013 (2017).
\bibitem{Min2014a} B. Min, S. D. Yi, K.-M. Lee, and K.-I. Goh, Phys. Rev. E {\bf 89}, 042811 (2014).
\bibitem{Hu2013} Y. Hu, D. Zhou, R. Zhang, Z. Han, C. Rozenblat, S. Havlin, Phys. Rev. E {\bf 88}, 052805 (2013).
\bibitem{Min2015} B. Min, S. Lee, K.-M. Lee, and K.-I. Goh, Chaos Soliton Fractals, {\bf 72}, 49 (2015).
\bibitem{Min2014b} B. Min, and K.-I. Goh, Phys. Rev. E {\bf 89}, 040802(R) (2014).
\bibitem{Baxter2014} G. J. Baxter, S. N. Dorogovtsev, J. F. F. Mendes, and D. Cellai, Phys. Rev. E {\bf 89}, 042801 (2014).
\bibitem{Brummitt2012} C. D. Brummit, K.-M. Lee, and K.-I. Goh, Phys. Rev. E {\bf 85}, 045102(R) (2012).
\bibitem{Lee2014} K.-M. Lee, C. D. Brummitt, and K.-I. Goh, Phys. Rev. E {\bf 90}, 062816 (2014).
\bibitem{Leicht2009} E. A. Leicht, and R. M. D'Souza, arXiv:0907.0894v1.
\bibitem{Lee2012} K.-M. Lee, J. Y. Kim, W.-k. Cho, K.-I. Goh, and I.-M. Kim, New. J. Phys. {\bf 14}, 033027 (2012).
\bibitem{Hackett2016} A. Hackett, D. Cellai, S. G\'omez, A. Arenas, and J. P. Gleeson, Phys. Rev. X {\bf 6}, 021002 (2016).
\bibitem{DeDomenico2014} M. De Domenico, A. Sol\'e-Ribalta, S. G\'omez, and A. Arenas, Proc. Natl. Acad. Sci. USA {\bf 111}, 8351 (2014).
\bibitem{Min2016} B. Min, S.-H. Gwak, N. Lee, and K.-I. Goh, Sci. Rep. {\bf 6}, 21392 (2016).
\bibitem{net_sci_book} A.-L. Barab\'asi, {\it Network science} (Cambridge University Press, Cambridge, 2016).
\bibitem{coauthorship} D. Lee, K.-I. Goh, B. Kahng, and D. Kim, Phys. Rev. E {\bf 82,} 026112 (2010).
\bibitem{callaway} D. S. Callaway, J. E. Hopcroft, J. M. Kleinberg, M. E. J. Newman, and S. H. Strogatz, Phys. Rev. E {\bf 64}, 041902 (2001).
\bibitem{pin1} R. V. Sol\'e, R. Pastor-Satorras, E. D. Smith, and T. Kepler, Adv. Complex Syst. {\bf 05,} 43 (2002).
\bibitem{pin2} A. V\'azquez, A. Flammini, A. Maritan, and A. Vespignani, ComPlexUs {\bf 1,} 38 (2003).
\bibitem{pin3} J. Kim, P. L. Krapivsky, B. Kahng, and S. Redner, Phys. Rev. E {\bf 66,} 055101 (2002).
\bibitem{rmp_2017} J. M. Kosteritz, Rev. Mod. Phys. {\bf 89,} 040501 (2017).
\bibitem{kt} J. M. Kosteritz, and D. J. Thouless, J. Phys. C. {\bf 5,} L124 (1972).
\bibitem{rmp_quantum} F. Duncan M. Haldane, Rev. Mod. Phys. {\bf 89,} 040502 (2017).
\bibitem{oh1} S. M. Oh, S.-W. Son, and B. Kahng, Phys. Rev. E {\bf 93}, 032316 (2016).
\bibitem{oh2} S. M. Oh, S.-W. Son, and B. Kahng (unpublished).
\bibitem{souza_nphy} R. M. D'Souza, and J. Nagler, Nat. Phys. {\bf 11}, 531 (2015).
\bibitem{nuno} N. Ara\'ujo, P. Grassberger, B. Kahng, K. J. Schrenk, and R. M. Ziff, {Eur. Phys. J.: Spec. Top.} {\bf 223}, 2307 (2014). 
\bibitem{jstat} D. Lee, Y. S. Cho, and B. Kahng, J. Stat. Mech. P124002 (2016).
\end{thebibliography}
\end{document}